\documentstyle[prl,aps]{revtex}
\begin{document}

\author{M. E. Portnoi(a) and I. Galbraith(b)}

\address{
(a) School of Physics, University of Exeter, Stocker Road, 
Exeter EX4 4QL, United Kingdom\\
(b) Physics Department, Heriot-Watt University, Edinburgh EH14 4AS, 
United Kingdom}

%\submitted{October 8, 2000}

\title{Exciton/free-carrier plasma in GaN-based quantum wells:\\
Scattering and screening}

\maketitle

\begin{abstract}
\noindent
The degree of ionisation of a two-dimensional electron-hole 
plasma is calculated in the low-density (Boltzmann) limit. 
The electron-hole interaction is considered for all states: 
optically active and inactive, bound and unbound. 
The theory is applied to exciton/free-carrier plasma in GaN-based 
quantum wells at room temperature.
\end{abstract}

%\pacs{71.35Ee, 73.20.Dx, 73.40Kp}

Room-temperature operation of blue quantum-well lasers, 
based on GaN\cite{Nichia} or ZnSe\cite{Sony}, is no longer a novelty.
However, the nature of lasing in wide-gap semiconductors and 
quantum wells has not yet been completely understood.
The large exciton binding energy in wide-gap semiconductors and 
quantum wells favours excitonic gain processes for 
which no satisfactory theoretical treatment exists\cite{Ian96}. 
A knowledge of the balance between excitons and free carriers 
is crucial in determining the dominant gain process. 
It is known that a naive application of the Mott criterion for 
the metal-insulator transition as well as the use of a single-bound-state 
law of mass action are insufficient, as screening of excitons by 
the electron-hole plasma and strong scattering of particles 
within the plasma both play a crucial role\cite{Zimm}.

For a consistent description of the electron-hole plasma 
at a temperature which is higher then the exciton binding energy,  
bound states and unbound scattering states should be treated on 
the same footing. In what follows we present such a consistent 
treatment for the purely two-dimensional (2D) case in the low-density 
(Boltzmann) limit.

Following an approach applied in 3D to nuclear matter\cite{Ropke},
an ionic plasma\cite{Plasma86}, and the electron-hole system in excited 
semiconductors\cite{Zimm}, we divide the total electron (hole) 
density between two terms:  
\begin{equation}
n_a~=~n_a^0~+~n_a^{corr}~.
\label{eq01}
\end{equation}
The first term $~n_a^0~$ is the density of uncorrelated quasiparticles 
with renormalized energies. Only this term should be taken for the 
screening radius calculation\cite{Zimm}.
All correlation effects both in the bound and continuum states are 
incorporated into the second term $~n_a^{corr}~$ which is called the 
correlated density. 
The lower index in Eq.~(\ref{eq01}) is a species index, 
$~a=e~$ for electrons and $~a=h~$ for holes.

In the low-density limit there is no need to go beyond two-particle 
correlations. This allows us to separate clearly the role of the
inter-particle Coulomb interaction from the phase-space filling effects.
It is tempting to relate $n_a^{corr}$ and  $n_a^{0}$ by a simple 
law of mass action with the single exciton bound state energy reduced 
by screening\cite{Ian96,Cing96}. The main shortcoming of this approach 
is a disregard of the strong scattering of unbound carriers. A complete 
account of scattering states as well as all (optically active and inactive)
bound states requires the calculation of a two-body partition function 
which involves summation over all two-particle states. 
In the low-density (non-degenerate) limit, for which there is no Pauli 
blocking, a 2D analogue of the modified mass action law reads    
\begin{equation}  
n_a^{corr}~=~\sum_b n_a^0 n_b^0 
{2\pi \hbar^2 \over \mu_{ab} k_B T} Z_{ab}~,
\label{eq02} 
\end{equation}
where $~\mu_{ab}=m_a m_b/(m_a+m_b)~$ is the reduced effective mass,
and $Z_{ab}$ is the two-body interaction part of the partition function.
Note that due to charge-neutrality the total electron-hole density 
$~n_e=n_h=n~$ is independent of species, whereas $~n_e^0 \neq n_h^0~$ and 
$~n_e^{corr} \neq n_h^{corr}~$ if the electron and hole have different 
masses.

The electron-hole part of the partition function which exhibits bound 
states (excitons) is given  by 
\begin{equation}
Z_{eh}=\sum_{m,\nu}\exp(-\beta E_{m,\nu}) 
  ~+~{1\over\pi} \int_0^\infty \left( \sum_{m=-\infty}^{\infty} 
{d\delta_m(k) \over dk} \right ) 
\exp\left(-\beta {\hbar^2 k^2 \over 2\mu_{eh}} \right)~dk~,
\label{eq03} 
\end {equation}
where $~\beta=1/(k_B T)$, $~m\hbar~$ is  the projection of the angular 
momentum onto the axis normal to the plane of 2D motion 
($m=0, \pm 1, \pm 2, ~\ldots$), 
$~\hbar^2 k^2/2\mu_{eh}~$ is the energy of the relative motion 
of the unbound (scattered) electron and hole, 
$~k~$ is the absolute value of the relative motion momentum,
$~\delta_{m}(k)~$ are the 2D scattering phase shifts introduced in the 
standard way\cite{SH67}, $~E_{m,\nu}~$ are the bound-state energies
(index $~\nu~$ enumerates bound states with given $~m$),
and the double sum in the first term ranges only over bound states. 
Equation (\ref{eq03}) is the 2D analogue of the Beth-Uhlenbeck 
formula\cite{BU37}, and it is derived\cite{PG99} in the same fashion 
as in the 3D case\cite{Landau5}.
The scattering term in the right-hand side of Eq.(\ref{eq03}) 
gives the contribution to $~Z_{eh}~$ of the electron-hole attraction 
in the continuum part of the energy spectrum.
The electron-electron and hole-hole parts of the partition function 
$~Z_{ee}~$ and $~Z_{hh}~$ contain the scattering term only.

Equations (\ref{eq01}-\ref{eq03}) provide a consistent description of the 
ionisation degree, defined as $~\alpha=n_e^0/(n_e^0+n_e^{corr})$. 
Technically the most difficult problem is to calculate the binding 
energies and scattering phase shifts in a screened Coulomb potential. 
We use for this purpose the variable-phase method\cite{VPA} 
known from scattering theory.  
In this method the scattering phase shift and the function 
defining bound-state energies can be obtained as a large 
distance limit of the phase function, which satisfies the 
first-order, nonlinear Riccati equation originating from the radial 
Schr\"{o}dinger equation.

In this paper we model the screened Coulomb interaction in a 2D plasma 
by the well-known Thomas-Fermi expression for a statically screened 
Coulomb potential\cite{SH67}:
\begin{equation}
V_s(\rho)=\mp {e^2 \over \epsilon} \int_{0}^{\infty}
{q J_0(q\rho) \over q+q_s}dq
=\mp {e^2 \over \epsilon} 
\left \{{1 \over \rho}-{\pi \over 2} q_s [{\bf H}_0(q_s \rho)  
- Y_0(q_s \rho)] \right\}~, 
\label{TF2D}
\end {equation} 
where $~q_s~$ is the 2D screening wavenumber (which depends on 
temperature and carrier density), $~\epsilon~$ is the static 
dielectric constant of the semiconductor,
$~J_0(x)$, $~Y_0(x)$, and $~{\bf H}_0(x)~$ are the Bessel functions 
of the first and of the second kind and the Struve function. 
The upper sign in Eq.~(\ref{TF2D}) is for electron-hole attraction,
the lower sign is for electron-electron or hole-hole repulsion.
Being the long-wavelength static limit of the random phase approximation
for a purely 2D case, Eq.~(\ref{TF2D}) is the simplest model for the screened 
Coulomb potential in 2D.   
Nevertheless, this expression reflects the fact that the statically 
screened potential in 2D decreases at large distances slower 
than in the 3D case.
Despite numerous realistic corrections, Eq.~(\ref{TF2D}) remains 
the most widely used approximation for the 2D screening, 
especially for the screened exciton problem. 
The variable-phase method application to scattering and bound states 
in the screened Coulomb potential (\ref{TF2D}) is described in detail 
in our recent paper\cite{PG97}. 
The method is especially effective for calculation of shallow-state
binding energies and low-energy scattering phase shifts.
Applying the variable phase method together with a 2D analogue of the 
Levinson's theorem\cite{Levinson}, we have found that with decreasing 
$~q_s~$ several bound states with different angular momenta appear 
simultaneously at certain integer values of $~1/(q_s a^*)$, where  
$~a^*~$ is the effective (excitonic) Bohr radius. This degeneracy is 
different from the well-known degeneracy of the unscreened 2D exciton 
states.
In the low-density limit ($q_s a^*\rightarrow0$) the number 
of bound states oscillates around $~1/(q_s a^*)~$ with the 
period and amplitude of oscillations proportional to $~1/\sqrt{q_s a^*}$.
Then, using the expression for the Thomas-Fermi 2D screening wavenumber 
$q_s$ for a two-component non-degenerate electron-hole plasma\cite{LSpM85}, 
$q_sa^*=4\pi~({\rm Ry}^*/k_BT)~(n_e^0a^{*2}+n_h^0a^{*2})$,
$~{\rm Ry}^*~$ being the excitonic Rydberg,
one can find the ratio of the free-carrier density to the total density.
For a model semiconductor with $m_e=m_h$, this 
ratio is equal to $2/3$ in the low-density limit.
This result is different from the single-bound-state law of the mass 
action, which gives a complete ionisation of bound states 
($\alpha\rightarrow1$) in the same limit.
 
Figure 1 shows the results from the calculation of the 
electron-hole part of the partition function, $~Z_{eh}$, which contains 
both the bound state sum and the scattering phase shift integral.
In this figure $~Z_{eh}~$ is plotted as a function of the inverse 
screening wave number $~1/q_s$ measured in units of the effective 
Bohr radius $~a^*$. The temperature is given in units of 
the bulk excitonic Rydberg $~{\rm Ry^*}$.
\begin{figure}
\includegraphics{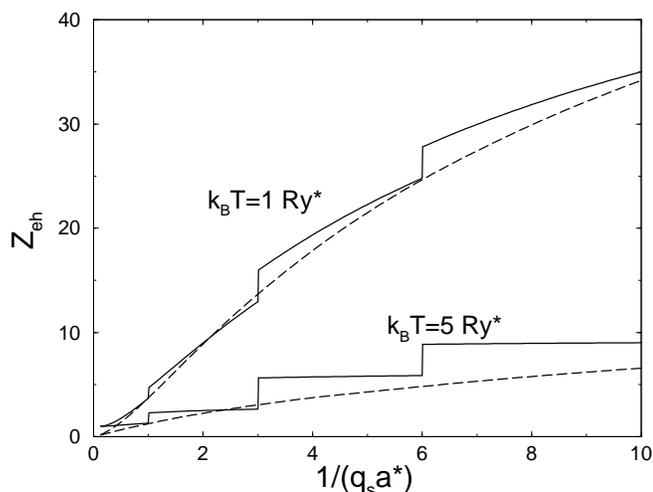}
\vskip 6.5truecm
\caption{    
The electron-hole part of the partition function, $~Z_{eh}~$ 
versus the screening length $~1/q_s~$ for two values of 
$~k_B T/{\rm Ry^*}$. Solid lines show the bound state contributions 
$~Z_{bound}~$ only. Dashed lines correspond to the total partition 
function with scattering states included.}
\end{figure}
To emphasize the role of scattering we show on the same plot the 
bound-state sum, $~Z_{bound}=\sum_{m,\nu} \exp(-\beta E_{m,\nu})$, 
which exhibits jumps whenever new bound states appear.
These jumps become higher with increasing screening length $~1/q_s~$ 
since several bound states appear simultaneously\cite{PG97}.
A proper account of scattering eliminates these unphysical jumps.

We calculated the density dependence of the degree of ionisation 
$~\alpha~$ in the 2D low-density (Boltzmann) limit for 
$~k_B T=1~{\rm Ry^*}$, which roughly corresponds 
to GaN at room temperature.  
Over a wide range of pair densities, $~0.01<n{a^*}^2<0.1$, 
$~\alpha~$ is almost independent of density, and it  
increases outside this range. 
The minimal value of the degree of ionisation, 
$~\alpha_{min} \approx 0.34$, corresponds to $~n{a^*}^2\approx0.04~$ 
($n\approx5\times 10^{12} \rm{cm^{-2}}~$ for GaN). 
Even at the relatively high densities of $10^{12}$ cm$^{-2}$  
we find that there is a single bound state having a binding energy 
of the order of $k_B T$ which is available to participate in 
excitonic lasing.

We also calculated the second virial coefficient of the dilute 2D 
electron-hole plasma in GaAs and GaN-based quantum wells at room
temperature. These calculations show a striking difference between
a strongly correlated exciton/free-carrier plasma in GaN and 
a nearly ideal electron-hole gas in GaAs.


\begin{thebibliography}{99}
\bibitem{Nichia} S. Nakamura {\it et al.}, Appl. Phys. Lett. 
{\bf 70}, 1417 (1997).

\bibitem{Sony} S. Taniguchi {\it et al.}, Electron. Lett., {\bf 32}, 
552 (1996).

\bibitem{Ian96} I. Galbraith, in: Microscopic Theory of Semiconductors: 
Quantum Kinetics, Confinement and Lasers, Ed. S.W. Koch, 
World Scientific, Singapore, 1996, (p. 211) and references therein.

\bibitem{Zimm} R. Zimmermann and H. Stolz, phys. stat. sol. (b) 
{\bf 131}, 151 (1985). 

\bibitem{Ropke} 
G. R\"{o}pke {\it et al.}, Phys. Rev. Lett. {\bf 80}, 3177 (1998).

\bibitem{Plasma86} W.-D. Kraeft, D. Kremp, W. Ebeling, and G. R\"{o}pke,
Quantum Statistics of Charged Particle Systems, Akademie-Verlag, 
Berlin, 1986.

\bibitem{Cing96} R. Cingolani {\it et al.}, Phys. Rev. B {\bf 54}, 17812 (1996).

\bibitem{SH67} F. Stern and W. E. Howard, Phys. Rev. {\bf 163}, 816 (1967).
 
\bibitem{BU37}  E. Beth and G.E. Uhlenbeck, Physica (Amsterdam) {\bf 4}, 
915 (1937).

\bibitem{PG99} M. E. Portnoi and I. Galbraith, Phys. Rev. B {\bf 60}, 5570 
(1999).

\bibitem{Landau5} L. D. Landau and E. M. Lifshitz, Statistical Physics, Pt.1,
Pergamon, New York 1980 (p. 236).

\bibitem{VPA} F. Calogero, Variable Phase Approach to Potential 
Scattering, Academic Press, New York 1967.

\bibitem{PG97} M.E. Portnoi and I. Galbraith, 
Solid State Commun. {\bf 103}, 325 (1997).

\bibitem{Levinson} N. Levinson, K. Dan. Vidensk. Selsk. Mat. Fys. Medd. 
{\bf 25}, 3 (1949).

\bibitem{LSpM85} J. Lee, H. N. Spector, and P. Melman,
J. Appl. Phys. {\bf 58}, 1893 (1985).
\end{thebibliography}
\end{document}